\documentclass[preprint,authoryear,3p,times,12pt,sort&compress]{elsarticle}
\usepackage{booktabs}
\usepackage{bm}
\usepackage{amsmath}
\usepackage{amsfonts}
\usepackage{amssymb}
\usepackage{graphicx}
\usepackage{fancyhdr}
\usepackage{gensymb}
\usepackage{fixltx2e}
\usepackage{caption}
\usepackage{textcomp}
\usepackage[colorlinks,linkcolor=blue,anchorcolor=green,citecolor=blue]{hyperref
}

\journal{International Journal of Solids and Structures}

\begin{document}

  \begin{frontmatter}

\title{A phase-field model of relaxor ferroelectrics based on random field 
theory}

\author[rvt]{Shuai Wang\corref{cor1}}
\ead{wang@mfm.tu-darmstadt.de} 

\author[rvt]{Min Yi\corref{cor1}}
\ead{yi@mfm.tu-darmstadt.de} 

\author[rvt]{Bai-Xiang Xu\corref{cor1}}
\ead{xu@mechanik.tu-darmstadt.de} 

\cortext[cor1]{Corresponding author}

\address[rvt]{Mechanics of Functional Materials Division, Institute of Material 
Science,
Technische Universit\"at Darmstadt, Jovanka-Bontschits-Strasse 2, Darmstadt 
64287, Germany}


\begin{abstract}
A mechanically coupled phase-field model is proposed for the first time to simulate the peculiar behavior of relaxor ferroelectrics. Based on the 
random field theory for relaxors, local random fields are 
introduced to characterize the effect of chemical disorder. 
This generic model is developed 
from a thermodynamic framework and the microforce theory and is implemented by 
a  
nonlinear finite element method. Simulation results show that the model can reproduce relaxor features, such as 
domain miniaturization, small  remnant polarization and large piezoelectric 
response. In particular, the influence of random field strength on these 
features are revealed.
Simulation results on domain structure and hysteresis behavior are discussed 
and compared with related experimental results.
\end{abstract}


\end{frontmatter}

  \section{Introduction}
In contrast with conventional perovskite structured ferroelectrics, relaxor 
ferroelectrics, also referred to as relaxors, exhibit high 
permittivity~\citep{park1997ultrahigh}, 
less hysteresis effect~\citep{shrout1990relaxor} and large piezoelectric 
response~\citep{cross1980electrostrictive}. 
The peculiar behavior of relaxors results from the abnormal domain structures, 
in which the short-range order dominates. 
The origin for the short-range order remains controversial. Besides the concept 
of polar
nanoregions~\citep{bokov2007recent}, the random 
field
model~\citep{Westphal1992,GFmodel} and the spherical random bond-random field
model~\citep{pirc1999spherical} have been used to explain the properties of
relaxors. The random field model was firstly proposed by
~\citet{Westphal1992}, 
referring to the original idea of
~\citet{imry1975random}. It was claimed that the relaxor behavior of 
Pb(Mg\textsubscript{1/3}Nb\textsubscript{2/3}
)O\textsubscript{3} (PMN) is due to the quenched random
electric fields.
These quenched fields can originate from
charged compositional fluctuations. 
In fact, most of the relaxors with the perovskite structure have heterovalent 
substitutions on 
the A- and/or B-site. For instance, in $AB_{1}B_{2}O_{3}$ structure, the 
differences in the charge and size of 
$B_{1}$ and $B_{2}$ cations make the structure asymmetric. When cooling down 
from 
high temperature, the concentrations of $B_{1}$ and $B_{2}$ cation 
tend to fluctuate and form chemical disorder. This can give rises to random 
internal electric field~\citep{Westphal1992,GFmodel}. 
The experiment
in linear birefringence supported this theory~\citep{Westphal1992,GFmodel}.
  
Nevertheless the exact mechanism, i.e., how the random field influences the 
domain structure in relaxors, remains unclear. There are very few theoretical 
attempts in the literature. 
For instance, Blinc and Pirc assumed Gaussian random field 
in 
their spherical random-bond-random-field 
model~\citep{blinc1999local,pirc1999spherical}, and showed analytically that 
these random fields can prohibit the phase 
transition from paraelectric phase into ferroelectric one. Instead, a 
long-range disordered phase forms. Using the same model, the authors also 
explained the dielectric non-linearity in 
PMN and predict 
local polarization distribution. Since the model is based on the Landau 
theory~\citep{landau1935theory} with homogeneous domain assumption, it has no 
access to the domain structure. More recently, ~\cite{ma2015lattice} 
proposed a lattice-based Ginzburg-Landau-type Hamiltonian and performed 
Monte-Carlo simulations on relaxors, by considering the random field theory. 
They found that the higher the random field is, the lower the freezing 
temperature becomes. The influence of the random field on the hysteresis and 
domain size were also studied. For the best knowledge of authors, there is no 
model for the study of domain structure on a larger scale and of the 
electromechanical coupling effect in relaxors.     
 
Phase-field approaches have been successfully applied to model the polarization 
switching in conventional 
ferroelectrics~\citep{kamlah1999phenomenological,chen2002phase, 
wang2004phase,landis2004non,xiao2005depletion,schrade2007domain,
kontsos2009computational,xu2010phase, 
munch2011domain,liu2013phase,su2015phase}
In particular, \cite{su2007continuum} proposed a 
phase-field model for 
ferroelectrics based on the microforce theory~\citep{gurtin1996generalized}. 
Thereby the total polarization is treated as the order parameter. Recently, 
\cite{keip2015coordinate} extend the 
phase-field~\citep{schrade2014invariant,schrade2015phase} simulation of 
ferroelectric single crystal into composites and polycrystals by homogenization. 
These models 
have become a useful tool in the study~\citep{wang2007phase} and 
design~\citep{liu2015giant} of domain structures in 
ferroelectrics. 
However, due to the complexity of relaxors, the phase-field theory has not been 
applied in this field. 

In this paper, we propose for the first time an electromechanical phase-field 
model for simulation of domain structures in relaxors. A continuum phase-field 
model is first established in the framework of thermodynamics and the 
microforce 
theory. Particularly, static random field is introduced, and its influence on 
the electrostatic energy is considered.
As a result, the evolution equations of the polarization and the constitutive 
equations are modified accordingly. Following the random 
field theory~\citep{GFmodel}, a Gaussian distribution generator is employed to 
produce 
the local random electric field. The
Gaussian distribution is generated based on the Box-Muller 
formula~\citep{box1958note}. Similar to our previous phase-field model for 
ferroelectrics~\citep{xu2010fracture}, the spontaneous polarization, instead of 
the total polarization, is taken as the order parameter, and the domain wall 
energy as well as the domain wall thickness appears explicitly as parameters in 
the model. This phase-field model is implemented in the finite element method 
(FEM) for its merit of 
robustness and flexible boundary condition. The model and its implementation 
are 
used to analyze the influence of random field on the macroscopic response and 
domain evolution under mechanical and/or electrical stimulation. 

The phase-field 
model and FEM implementation are elaborated in Section 2 and 3, respectively, 
whereas Section 4 presents the simulation results based on the model, in order 
to reveal the influence of random field. In Section 4.1 the dependency of the 
equilibrium domain structure on the variance of the random field distribution 
is 
first revealed by numerical results. The domain structure evolution under 
bipolar electric field is demonstrated in Section 4.2, along with the 
dielectric 
and butterfly hysteresis. Section 4.3 is concerned with ferroelastic switching 
of the domains in relaxors. Finally the electromechancial loading is considered 
in Section 4.4. Results show that the model can predict characteristic features 
of relaxors, such as domain miniaturization, decreasing of  remnant
polarization 
and increased electromechanical effect.

\section{Phase-field model}
\subsection{Governing equations}
   Denote the space occupied by a relaxor ferroelectric body and its boundary 
by 
$\mathcal{B}$ and $\mathcal{\partial B}$, respectively. At each point in 
$\mathcal{B}$ the mechanical 
equilibrium 
\begin{equation}\label{blc1}
 \sigma_{ij,i}+f_j=0 \quad \text{in} \quad \mathcal{B}
\end{equation}
must be satisfied, where $\sigma_{ij}$ is the Cauchy stress in the Cartesian 
space, and $f_i$ is the body force in the $i$\textsuperscript{th} direction. 
Hereafter, 
the Einstein notation 
is implied, 
 the Latin symbols $i, j, k, l$ run from 1 to 3, and the comma in the 
subscript represents the spatial partial derivative. By assuming small 
deformation, the following strain measures are used
\begin{equation}\label{epsilon}
  \varepsilon_{ij} = \frac{1}{2} (u_{i,j}+u_{j,i}),
\end{equation}
where $u_i$ the displacement vector. 
For the point at the 
boundary surface, displacement or force boundary conditions can be applied,
\begin{subequations}
\begin{align}
  u_i = \bar{u}_i  \quad \text{on} \quad \mathcal{\partial{B}}_u,\\
  \sigma_{ij}n_i = \bar{t}_j \quad \text{on} \quad  
\mathcal{\partial{B_\sigma}}, 
\end{align}
\end{subequations}
where $n_i$ is the normal vector on the surface, and $t_j$ the traction vector. 
Moreover, $\bar{u}_i$ denotes the displacement prescribed on the surface part 
$\partial \mathcal{B}_u$, while $\bar{t}_j$ denotes the traction on the surface 
part $\partial \mathcal{B}_\sigma$.

The electric quantities such as electrical displacement $D_i$ and volume 
charge density $q$ are governed by the quasi-static equilibrium derived from 
Maxwell equations, i.e.,
\begin{equation}
 D_{i,i}=q  \quad \text{in} \quad \mathcal{B}.
\end{equation}
The electric field can be given as 
\begin{equation} \label{ephi}
 E_i=-\phi_{,i}.
\end{equation}
in which $\phi$ is the electric potential. 
The boundary conditions for the electric quantities can be given as
\begin{subequations}
\begin{align}
  \phi &= \bar{\phi}  \quad \text{on} \quad \mathcal{\partial{B}}_{\phi},\\
  D_{i}n_i &= -\bar{\omega} \quad \text{on} \quad \mathcal{\partial{B}}_D ,
\end{align}
\end{subequations}
where $\bar{\phi}$ is the surface charge density 
applied on the surface part 
$\partial \mathcal B_{\phi}$, and $\bar{\omega}$ is the surface charge density 
applied on the surface part $\partial \mathcal B_D$. 

The microforce theory, 
originally proposed by \cite{gurtin1996generalized}, is utilized for the 
derivation of the polarization evolution equation in the following subsection. 
Thereby the balance law of the microforces is introduced first. In the aimed 
phase field, the order parameter is chosen to the spontaneous polarization 
$P_i$. The 
microforce system associated with $P_i$ is characterized by 
a generalized stress tensor $\xi_{ij}$ together with related body forces $\pi_i$ 
and $\gamma_i$ that 
represent, respectively, the internal and external forces distributed over the 
volume $\mathcal{B}$. Given an arbitrary control volume $\mathcal R$ (subregion 
of $\mathcal B$), 
with 
$n_i$ the unit normal vector on the surface $\partial \mathcal R$, the 
microforce balance holds,
\begin{equation} \label{microint}
  \int_{\partial{\mathcal R}} \xi_{ji}n_j dS+\int_{\mathcal R} 
\pi_i
dv+\int_{\mathcal R}\gamma_i dv=0.
\end{equation}

By using the Gauss law and noting that Eq.~(\ref{microint}) should hold for 
any 
arbitrary volume, it yields
\begin{equation}\label{microforceblc}
   \xi_{ji,j}+ \pi_i+\gamma_i=0.
\end{equation}
On the boundary, there lays 
\begin{equation}\label{microbc}
   \xi_{ji}n_j=\bar \mu_i\quad \text{on} \quad   \partial \mathcal{R},
\end{equation}
where $\bar \mu_j$ is the surface microforce applied on the surface part 
$\partial \mathcal{R}$.
In the next section, we will use these microforces to derive 
the 
constitutive and the evolution equations.

\subsection{Constitutive and evolution equations}
In order to get the constitutive relations and evolution equation, 
thermodynamic relations are considered.
According to the second law of thermodynamics, under the isothermal 
condition, the change rate in the Helmholtz free energy in the control 
volume should not be greater than the external power expended on the control 
volume, i.e.,
\begin{equation}\label{noneq}
 \{  \int_\mathcal B\mathcal H dv \}^. \leq  \mathcal W^{ext}.
\end{equation}

The free energy $\mathcal 
H$ 
is a functional of $\varepsilon_{ij}, D_i, P_i,P_{i,j}$ and high 
order items of these 
parameters. To simplify the model and not lost generality, one can write 
$\mathcal H$ as
\begin{equation}
   \mathcal{H}=\mathcal{H}(\varepsilon_{ij}, D_i, P_i, P_{i,j}).
\end{equation}
Applying Taylor expansion and retaining only the first order term, the left 
hand of Eq.~(\ref{noneq}) can be written as
\begin{equation}\label{Hpartial}
   \{  \int_\mathcal B\mathcal H dv \}^. = (
   \int_\mathcal B \frac{\partial \mathcal H}{\partial \varepsilon_{ij} 
}\dot{\varepsilon}_{ij} 
+  \int_\mathcal B \frac{\partial \mathcal H}{\partial D_i}\dot{D}_i 
+  \int_\mathcal B \frac{\partial \mathcal H}{\partial P_i}\dot{P}_i 
+ \int_\mathcal B \frac{\partial \mathcal H}{\partial P_{i,j}}\dot{P}_{i,j}
) dv.
\end{equation}
$W^{ext}$ is the power done by external resources, and have components of 
surface power and body power, and corresponding 
power to $P_i$ 
can be present by microforce mentioned above,
\begin{equation}\label{Wext}
  W^{ext}=\int_\mathcal B (f_i\dot u_i+\phi \dot q+\gamma_i \dot P_i)dv+
         \int_\mathcal {\partial B}  (\bar{t}_i\dot u_i+\phi 
\dot {\bar{\omega}}+\bar{\mu}_i 
\dot 
P_i) ds.
\end{equation}
What should be noticed is that the internal force $\pi_i$ do 
not contribute to the external power $W^{ext}$. After the Legendre 
transformation, the electric enthalpy $\mathcal H_2(\varepsilon_{ij}, E_i, 
P_i,P_{i,j})$ can be derived as $\mathcal H_2= \mathcal H-E_iD_i$. Inserting 
Eq.~(\ref{Hpartial}) 
Eq.~(\ref{Wext}) and Eq.~(\ref{microforceblc}) into Eq.~(\ref{noneq}) 
and applying the divergence theorem, one arrives at
\begin{equation}\label{thermo2}
\begin{split}
 \int_\mathcal B \{
(\sigma_{ji}-\frac{\partial \mathcal H_2}{\partial {\varepsilon_{ij}}} 
)\dot{\varepsilon }_{ij}-
(D_i+\frac{\partial \mathcal H_2}{\partial E_i})\dot{E}_i-
(\frac{\partial \mathcal H_2}{\partial P_i}+\pi_i)\dot{P}_i+ \\
(\xi_{ji}-\frac{\partial \mathcal H_2}{\partial P_{i,j}})\dot{P}_{i,j}
 \}dv
\geq 0.
\end{split}
\end{equation}
From Eq.~(\ref{thermo2}) one can obtain three constitution relations 
\begin{equation}\label{constitution}
\sigma_{ji}=\frac{\partial \mathcal H_2}{\partial {\varepsilon_{ij}}} 
,\quad
D_i=-\frac{\partial \mathcal H_2}{\partial E_i},\quad
\xi_{ji}=\frac{\partial \mathcal H_2}{\partial P_{i,j}},
\end{equation}
and one inequality 
\begin{equation}
 (\frac{\partial \mathcal H_2}{\partial P_i}+\pi_i)\dot{P}_i \leq 0
\end{equation}
due to dissipative 
process.

To ensure the inequality one can assume that 
\begin{equation}\label{governing1}
 \frac{\partial \mathcal H_2}{\partial P_i}+\pi_i=-\beta \dot{P_i},
\end{equation}
where the non-negative constant $\beta$ indicates the mobility parameter. 
By applying the balance equation of microforce Eq.~(\ref{microforceblc}), one 
can 
rewrite Eq.~(\ref{governing1}) in the
following form
\begin{equation}\label{governing2}
\beta \dot{P}_i =-\frac{\partial \mathcal{H}_2}{\partial P_i}+
\xi_{ji,j}+
\gamma_i = -\frac{\partial \mathcal{H}_2}{\partial P_i}+
\left(\frac{\partial \mathcal H_2}{\partial P_{i,j}}\right)_{,j}+
\gamma_i.
\end{equation}

\subsection{Electrical enthalpy}\label{enthalpy}

In our model, the electrical enthalpy consists of five parts, i.e.,
\begin{equation}\label{sumH}
\mathcal{H}_2=\mathcal{H}^{ela}+\mathcal{H}^{ele}+\mathcal{H}^{coup}
+ \mathcal{H}^{sep}+\mathcal{H}^{grad},
\end{equation}
in which $\mathcal{H}^{ela}$, $\mathcal{H}^{ele}$, $\mathcal{H}^{coup}$, 
$\mathcal{H}^{sep}$ and $\mathcal{H}^{grad}$ represent elastic energy density, 
electrical energy density, electromechanical coupling energy density, domain 
separation energy density and interface energy density, 
respectively. The domain separation energy density resembles the 
Landau-Devonshire free energy, and the interface energy density characterizes 
to 
the energy stored in the domain wall by a function of the 
gradient of the order parameter. These energy terms take the follow specific 
form,
\begin{equation}\label{Hcom}
\left\{
 \begin{aligned}
 \mathcal{H}^{ela}&=\frac{1}{2} c_{ijkl} \varepsilon_{ij}^{e} 
\varepsilon_{kl}^{e}\\
 \mathcal{H}^{ele}&=-\frac{1}{2} k_{ij} E_i E_j-P_iE_i\\
 \mathcal{H}^{coup}&=-b_{ijk} \varepsilon_{ij} E_k\\
\mathcal{H}^{sep}&=\beta_1(G,\lambda)\psi(P_i)\\
\mathcal{H}^{grad}&=\beta_2(G,\lambda) P_{i,j} P_{i,j}.
\end{aligned}
\right.
\end{equation}
The quantities induced in the last equations are explained in the following.
Firstly, $\varepsilon_{ij}^{e}$ stands for the elastic strain, which is further 
given by the difference between the mechanical strain $\varepsilon_{ij}$ and 
the spontaneous strain $\varepsilon^0_{ij}$, i.e., 
$\varepsilon_{ij}^{e}=(\varepsilon_{ij}-\varepsilon^0_{ij}(\boldsymbol 
P))$. The spontaneous strain 
$\varepsilon^0_{ij}(\boldsymbol 
P)$ is caused by the spontaneous polarization. According to 
\cite{huo1997modeling},
\begin{equation}
\varepsilon^0_{ij}(\boldsymbol P)=\frac{3}{2}\varepsilon_{sat} 
\frac{\sqrt{P_iP_i}}{P_{0}}(n_in_j-\frac{1}{3}\delta_{ij}),
\end{equation}
where $\delta_{ij}$ is the Kronecker delta, $n_i$  the unit vector of 
$\boldsymbol P$, $\varepsilon_{sat}$  the maximum  remnant
strain, and $P_{sat}$ the maximum  remnant polarization.
The stiffness tensor $c_{ijkl}$ and the permittivity tensor $k_{ij}$ are 
assumed 
to be independent of the polarization. For the piezoelectric tensor $b_{ijk}$, 
we used the 
representation~\citep{kamlah2001ferroelectric}
\begin{equation}
\begin{split}
 b_{kij}(\boldsymbol P)=&  \frac{\sqrt{P_iP_i}}{P_{0}} \Big\{
d_{33}n_in_jn_k+d_{31}(\delta_{ij}-n_in_j)n_k \\
 &+\frac{1}{2} d_{15}\left[  (\delta_{ki}-n_kn_i)n_j+(\delta_{kj}-n_kn_j)n_i
\right]
\Big\},
\end{split}
\end{equation}
in which $d_{33}, d_{31}, d_{15}$ are linear piezoelectric constants at 
the
poled state. The coefficient functions $\beta_{1}$ and $\beta_{2}$ are  
related to the domain wall energy $G$ and the domain wall thickness 
$\lambda$~\citep{xu2010phase}. The function $\psi(P_i)$ is a Landau-Devonshire 
free energy function with potential wells. The 
2-dimensional form of $\psi$ is shown in section \ref{2d}.

It should be noted that the electric field has three contributions
\begin{equation}
E_i=E_i^{d}+E_i^{ext}+E_i^{random},
\end{equation}
where $E_i^{d}$ represents the depolarization field, 
$E_i^{ext}$  
the applied electric field, and $E_i^{random}$  the random field induced by 
chemical disorder. In the model, 
${E}_i^{random}$ is assumed to be static and obeys the Gaussian 
distribution $\mathcal{N}(0,\Delta)$. The expectation of the random field 
Gaussian distribution is set to be zero, while the variance of the distribution 
is denoted by $\Delta$. The 
numerical generation is explained in section \ref{2d}. The two contributions 
$E_i^{d}$ and  $E_i^{ext}$ are calculated from the 
electric potential $\phi$ by Eq.~(\ref{ephi}). To be in accordance with 
Eq.~(\ref{ephi}), 
the sum of $E^{d}$ and $E^{ext}$ is denoted as $E^{\phi}$. It reveals
\begin{equation}\label{random}
\left\{
 \begin{aligned}
 &E_i=E^{\phi}_i+E^{random}_i
 \\
 &E^{\phi}_i=\phi_{,i}=E_i^{d}+E_i^{ext}
 \\
&E_i^{random}=\mathcal{N}(\mu, \Delta).
 \end{aligned}
\right.
\end{equation}

Substituting Eqs.~(\ref{random}) into Eqs.~(\ref{Hcom}) and then 
(\ref{sumH}), one 
gets the expression of $\mathcal{H}_2$,
\begin{equation}
\begin{aligned}
\mathcal H_2=& \frac{1}{2} c_{ijkl} \varepsilon_{ij}^{e} 
\varepsilon_{kl}^{e} -\frac{1}{2} 
k_{ij}(E_i^{\phi}+E_i^{random})(E_j^{\phi}+E_j^{random})\\
&-P_i(E_i^{\phi}+E_i^{random})-b_{ijk} 
\varepsilon_{ij}(E_k^{\phi}+E_k^{random})\\
&+\beta_1(G,\lambda)\psi(P_i)+\beta_2(G,\lambda) P_{i,j} P_{i,j}.
\end{aligned}
\end{equation}
Inserting the expression into Eqs.~(\ref{constitution}) and (\ref{governing2}) 
and assuming $\gamma_i = 0$, the constitutive equations and the evolution 
equation can be rewritten as, 
\begin{equation}\label{constitutive}
\left\{
 \begin{aligned}
&\sigma_{ij}=c_{ijkl}\epsilon_{kl}^{e}-b_{jik}E_{k}^{\phi}-b_{jik}E_{k}^{random}
 \\
&D_i=k_{ij}E_j^{\phi}+k_{ij}E_j^{random}+P_i+b_{ijk}\epsilon_{jk}
 \\
&\beta 
\dot{P_i}=E_i^{\phi}+E_i^{random}-\beta_{1}\frac{\partial{\psi}}{\partial{P_i}}
-\beta_{2}\frac{\partial{P_{k,l}P_{k,l}}}{\partial{P_i}}+2\beta_{2}(P_{i,j})_{,j
}.
 \end{aligned}
\right.
\end{equation}

\section{Numerical implementation}\label{num}
\subsection{Weak form}
The governing equations for the phase-field model are 
Eq.~(\ref{blc1})-Eq.~(\ref{microbc}), 
Eq.~(\ref{constitutive}). To obtain solutions to these 
equations, the finite element method is adopted. Mechanical displacement, 
electric potential and 
spontaneous polarization are chosen as the degrees 
of freedom, by following the choice of electrical enthalpy as the energy 
type. The weak form of the field equation can be given as
\begin{equation}\label{weak1}
\begin{split}
 \int_\mathcal B\beta \dot P_i\delta P_i 
dv+ 
\int_\mathcal B(\sigma_{ji}\delta\varepsilon_{ij}-D_i\delta E_i+\eta_i\delta 
P_i+\xi_{ji}\delta P_{i,j} )dv= \\
\int_\mathcal B (f_i\delta u_i -q\delta\phi+\gamma_i\delta P_i)dv+\int_{
\partial \mathcal B} (\bar{t}_i\delta u_i -\bar \omega \delta \phi +\bar{\mu}_i 
\delta P_i)ds.
\end{split}
\end{equation}
The left-hand of the equation represents the generalized virtual strain energy. 
The first part of the right-hand stands for the virtual work done in the body, 
and the second part for the virtual work done on 
the boundary. The stresses, electric displacements and microforces can be 
derived 
from Eqs.~(\ref{constitutive}). 

By considering that $u_i, E_i ,P_i$ are the independent variables, one obtains 
from 
Eq.~(\ref{weak1}) the following weak forms,
\begin{equation}\label{weak2}
\left\{
\begin{aligned}
 -\mathcal 
\int_\mathcal B \sigma_{ji}\delta\varepsilon_{ij} dv+
\int_\mathcal B f_i\delta u_i dv+\int_{
\partial \mathcal B} \bar{t}_i\delta u_i ds&=0 \\
-\int_\mathcal B   D_i\delta E_i  dv+ 
\int_\mathcal B  q\delta\phi dv+\int_{
\partial \mathcal B} \bar \omega \delta \phi ds &=0  \\
 -\int_\mathcal B\beta \dot P_i\delta P_i 
dv- 
\int_\mathcal B (\frac{\partial \mathcal{H}}{\partial P_i}\delta P
_i+\frac{\partial \mathcal{H}}{\partial P_{i,j}}\delta P_{i,j} )dv +
\int_{
\partial \mathcal B} \mu_i 
\delta P_i ds&=0.
\end{aligned}
\right.
\end{equation}

\subsection{2-Dimensional case}\label{2d}
The weak form of the field equation is suitable for 3-dimensional cases. 
However, it is adequate to 
explore the domain evolution in 2-dimensional, since in the present paper merely 
qualitative study of the relaxor behavior is intended. Meanwhile, this saves 
calculation time. 
It is non-trivial to reduce the 3-dimensional rhombohedral system into a
2-dimensional one, while the tetragonal and the orthorhombic system can be 
represented in 
two dimensions by limiting the allowed polarization directions to one principal
plane. For tetragonal system, polarization has six possible orientations. In 
2-dimensional case, we focus on one plane with 4 possible 
orientations, through which both 90\degree\  and 180\degree\  domain switching 
 can be presented.

The Landau-Devonshire free energy function $\psi$ in 2-dimensional case has a 
simple 
form. Here we only chose parts of the Taylor polynomial and expand $\psi$ to the
sixth order,
\begin{equation}
\begin{split}
 \psi=a_1+a_2(P_1^2+P_2^2)+a_3(P_1^4+P_2^4)+a_4(P_1^2P_2^2)+a_5(P_1^6+P_2^6).
\end{split}
\end{equation}
The parameters $a_1$ to $a_5$ are chosen properly to allow 
$90^{\circ}$and $180^{\circ}$ domain switching. Meanwhile $\psi$ 
must guarantee that it takes local minimum at ($\pm P_0$,0) and 
(0,$\pm P_0$). The five parameters used in the simulations are derived from the 
coefficients published in 
\cite{bell1984phenomenological}.

The discretization of 2-dimensional model is achieved by four-node linear 
elements. The spontaneous polarization \underline{\textbf{P}}($P_1, P_2$), the 
electric potential $\phi$, and mechanical displacement 
\underline{\textbf{u}}($u_1, u_2$) are taken as independent variables. 
Therefore, each node has five degrees of freedom 
\[\textbf{\underline{d}}\textsuperscript{I}={[P_1^I, P_2^I, \phi^I, u_1^I, 
u_2^I]}^T,\] in which the superscript $I$ indicates the element nodal number 
and 
the underlined bold symbol denotes a matrix. 
In this section, matrix  
instead of index notation is used to clearly shows the components of the tensor.
A Voigt notation is used in the 
discretized equations. Detailed components of the matrix can be found in the 
Appendix.
By introducing the shape functions, the values within the element can be 
expressed by nodal values,
\begin{equation}
\left\{
\begin{aligned}
                {\underline{\bm{u}}}=&\sum_{I}N_u^I\underline{\bm u}^I;
  \ \quad   {{\delta \underline {\bm{u}}}}=\sum_{I}N_u^I{\delta \underline {
\bm{u}}}^I;\\
  {{\phi}}=&\sum_{I}N_{\phi}^I{\phi}^I;
  \ \quad   {{\delta {\phi}}}=\sum_{I}N_{\phi}^I{\delta {\phi}}^I;\\
  {\underline{\bm P}}=&\sum_{I}N_P^I\underline {\bm P}^I;
  \quad   {{\delta \underline {\bm P}}}=\sum_{I}N_P^I{\delta  \underline {\bm 
P}}^I;\\
  {\underline{\dot {\bm P}}}=&\sum_{I}N_P^I\underline{\dot{{\bm P}}}^I.
 \end{aligned}
 \right.
\end{equation}

The strain, the electric field and the polarization gradient can be calculated 
by introducing corresponding derivative matrices, and then the stress and the 
electric displacement can be obtained through Eqs.~({\ref{constitutive}}),
\begin{equation}
\left\{
\begin{aligned}
   & {\underline{\bm \varepsilon}}=\sum_I\underline {\bm B}_u^I\underline 
u^I;\quad
{\underline{\bm 
\varepsilon}^e}={\underline{\bm \varepsilon}}-{\underline{\bm \varepsilon } 
^0(\boldsymbol{P})};\quad
      {\underline{\boldsymbol{E}}}=-\sum_I\underline {\bm 
B}_{\phi}^I{\phi}^I;\quad
      {\nabla{\underline{\boldsymbol P}}}=\sum_I\underline {\bm 
B}_{\nabla}^I\underline{\bm P}^I;\\
   &    \underline{\boldsymbol{\sigma}}=\underline{\boldsymbol 
C}{\underline{\boldsymbol \varepsilon}}-{\underline{\boldsymbol 
b}}^T{\underline{\boldsymbol E}};\quad
      {\underline{\boldsymbol{D}}}={\underline{\boldsymbol 
b}}\,{\underline{\boldsymbol \varepsilon}}+
      {\underline{\boldsymbol k}}{\underline{\boldsymbol 
E}}+{\underline{\boldsymbol P}}.
       \end{aligned}
 \right.
\end{equation}
From Eqs.~(\ref{weak2}), the element residuals can be given by the integration 
over the element area $\mathcal B^e$:
\begin{equation} \label{residual}
 \left\{
 \begin{aligned}
  {\underline{\boldsymbol{R}}}^I_{u}=&-\int_{\mathcal B^e} {\underline {\bm 
B}^{I}_{u}}^{T}{\underline{\boldsymbol{\sigma }}}dv
  \\
  R^I_{\phi}=&-\int_{\mathcal B^e} {\underline {\bm B}^{I}_{\phi}}^{T} 
{\underline{\boldsymbol{D}}}  dv
  \\
  {\underline{\boldsymbol{R}}}^I_{P}=&-\int_{\mathcal B^e}
   \{
   {\beta} N^I_{P}\underline{\dot{{\boldsymbol{P}}}}+
   N^I_{P}\frac{\partial \mathcal{H}}{\partial 
{\underline{\boldsymbol{P}}}}+{\underline {\bm B}_{\nabla}^I}^{T}\frac{\partial 
\mathcal H}{\partial 
{{\nabla{\underline{\boldsymbol P}}}}}
   \}
  dv,
 \end{aligned}
\right.
\end{equation}
where
\begin{equation}
\left\{
 \begin{aligned}
\frac{\partial \mathcal{H}}{\partial 
{\underline{\boldsymbol{P}}}}=&
-\frac{{\partial \underline{\bm \varepsilon^0}}^T}{\partial 
\underline{\boldsymbol{P}}}{\underline{\boldsymbol{\sigma}}}
-({\underline {\bm \varepsilon}^0}^T
\frac{\partial {\underline{\boldsymbol{b}}}^T}{\partial 
\underline{\boldsymbol{P}}}
+\underline{\boldsymbol{1}})
(\underline {\bm E}^{\phi}+\underline {\bm E}^{random})
+\beta_{1}\frac{\partial \psi}{ {\partial \underline {\boldsymbol{P}}}}\\
\frac{\partial \mathcal{H}}{\partial 
{\nabla \underline{\boldsymbol{P}}}}=&
\beta_2 \nabla \underline{\boldsymbol{P}}.
 \end{aligned}
\right.
\end{equation}
The random field ${\underline 
{\bm E}}^{random}$ has two independent components, which are subject to the same 
Gaussian distribution. According to \cite{box1958note}, the following formula is 
used  
\begin{equation}
 \left\{
 \begin{aligned}
  E_x^{random}=&\Delta^2(-2\log U_1)^{\frac{1}{2}}\cos 2\pi U_2+\mu\\
  E_y^{random}=&\Delta^2(-2\log U_1)^{\frac{1}{2}}\sin 2\pi U_2+\mu.
 \end{aligned}
  \right.
\end{equation}
Hereby $U_1, U_2$ are two independent random numbers which obeys the 
standard uniform distribution, while $\Delta$ and $\mu$ govern the 
variance and the expectation of the Gaussian distribution. The stiffness matrix 
can be calculated 
thereafter by
\begin{equation}
\left\{
\begin{aligned}
 {\underline {\boldsymbol{K}}}^{IJ}_{uu}=-\frac{\partial{\underline 
{\boldsymbol{R}}}^I_{
u}}{\partial{\underline {\bm u}^J}};\quad
 {\underline {\boldsymbol{K}}}^{IJ}_{u\phi}=-\frac{\partial{\underline 
{\boldsymbol{R}}}^I_{
u}}{\partial{ {\phi}^J}};\quad
 {\underline {\boldsymbol{K}}}^{IJ}_{uP}=-\frac{\partial{\underline 
{\boldsymbol{R}}}^I_{
u}}{\partial{\underline {\bm P}^J}};\\
 {\underline {\boldsymbol{K}}}^{IJ}_{\phi u}=-\frac{\partial{\underline 
{\boldsymbol{R}}}^I_{
\phi}}{\partial{\underline {\bm u}^J}};\quad
 {\underline {\boldsymbol{K}}}^{IJ}_{\phi \phi}=-\frac{\partial{\underline 
{\boldsymbol{R}}}^I_{
\phi}}{\partial{{\phi}^J}};\quad
 {\underline {\boldsymbol{K}}}^{IJ}_{\phi P}=-\frac{\partial{\underline 
{\boldsymbol{R}}}^I_{
\phi}}{\partial{\underline {\bm P}^J}};\\
 {\underline {\boldsymbol{K}}}^{IJ}_{Pu}=-\frac{\partial{\underline 
{\boldsymbol{R}}}^I_{
P}}{\partial{\underline {\bm u}^J}};\quad
 {\underline {\boldsymbol{K}}}^{IJ}_{P\phi}=-\frac{\partial{\underline 
{\boldsymbol{R}}}^I_{
P}}{\partial{{\phi}^J}};\quad
 {\underline {\boldsymbol{K}}}^{IJ}_{PP}=-\frac{\partial{\underline 
{\boldsymbol{R}}}^I_{
P}}{\partial{\underline {\bm P}^J}}.
\end{aligned}
\right.
\end{equation}

For the only time-dependent term $\dot{P}_i$, implicit backward Euler  time 
integration method can be 
adopted. The non-zero damping matrix $\underline{\boldsymbol{D}}_{PP}^{IJ}$ can 
be calculated by
\begin{equation}
\left\{
 \begin{aligned}
\dot{\underline{\boldsymbol{P}}}=&\frac{\underline{\boldsymbol{P}}^{t+\tau}
-\underline{\boldsymbol{P}}^{t}}{\tau}\\ 
\underline{\boldsymbol{D}}_{PP}^{IJ}=&-\frac{\partial{\underline{\boldsymbol{R}}
_P^I}}{\partial{\dot{\underline{\boldsymbol{P}}^J}}},
 \end{aligned}
  \right.
\end{equation}
where $\tau$ is the time step, and $t$ the current time. With stiffness and 
damping matrix, the element iteration matrix $\underline{\boldsymbol{S}}$ can be 
assembled, in which $ 
\underline{\boldsymbol{S}}_{PP}^{IJ}=\underline{\boldsymbol{K}}_{PP}^{IJ}+\frac{
1}{\tau}\underline{\boldsymbol{D}}_{PP}^{IJ}$.
For a given sets of nodal values at time $t$, the non-linear algebraic 
equations can be solved by the Newton-Raphson method. The model is implemented 
as a user element in the software FEAP~\citep{feap}. 

\section{Simulation results and discussion}
Four cases are simulated based on the relaxor 
model presented above. 
Firstly, the equilibrium domain configurations of a relaxor single crystal with 
different random fields are simulated. 
Secondly, the ferroelectric response under bipolar loading is analyzed. Domain 
evolution with 
various random fields shows different features. 
Thirdly, pure mechanical loading is applied, and the influence of random field 
is demonstrated.
Finally, domain evolution under electromechanical loading is discussed. In all 
the simulations, the 
 body force $f_i$ and charge $q$ are neglected. The parameters used in the 
simulations are listed in Table 1.
  
\subsection{Equilibrium state of relaxor single crystal}
Relaxation process of relaxors is firstly simulated 
based on this model. The size of the simulated 
sample is 100 $\times$ 100 nm\textsuperscript{2}, 
with mesh size of 1 nm. All the boundaries are traction-free and 
charge-free. Initial polarization distribution is assumed to be random.

\begin{figure}[!hbp]
\centering
\includegraphics[width=\textwidth]{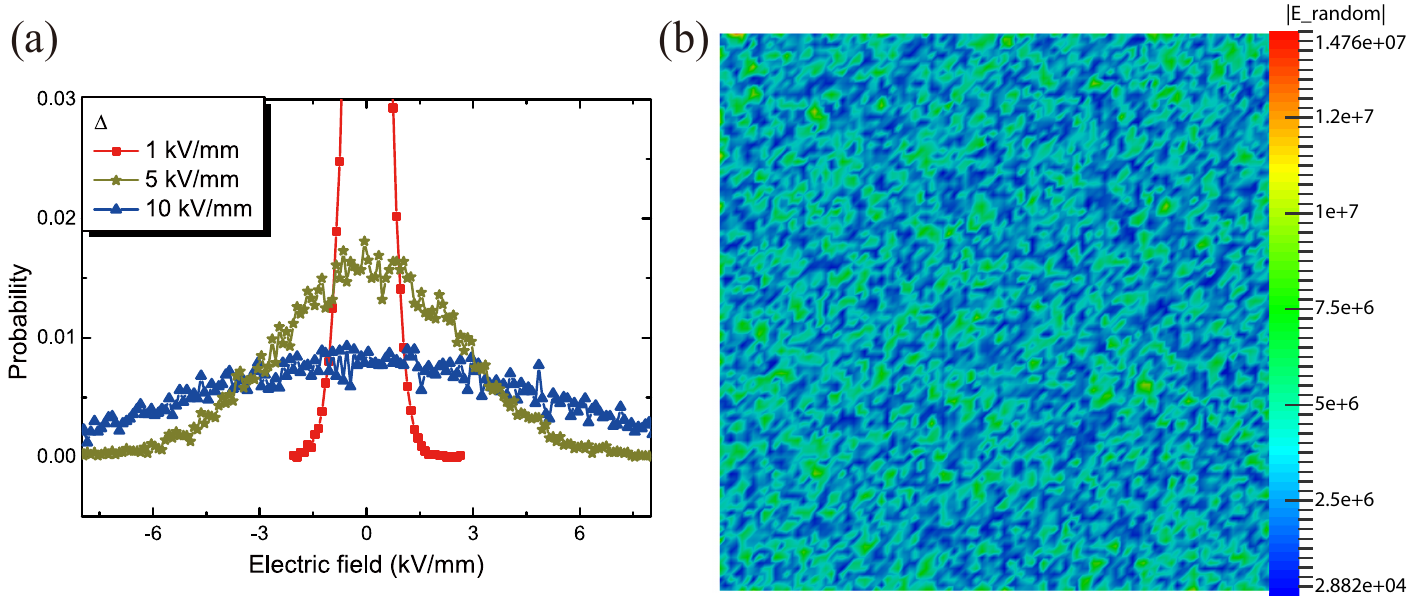}
\caption{ (a) Probability diagram of the local random field distributions 
with 
different $\Delta$. (b) Random field distribution within the sample for the 
case 
of $\Delta$= 5 kV/mm. Legend unit: 
V/m.}
\label{fig1}
\end{figure}

One of the key factors which can influence the domain configuration is the 
local 
random field which obeys the Gaussian distribution $\mathcal N 
(0,\Delta)$. In the simulation, the mean value of the random fields is set to be 
zero, to ensure that there is no macroscopic bias field. Thus, the variance 
$\Delta$ of the Gaussian distribution 
accounts for the strength of the local random field. Three values of $\Delta$, 
i.e.,  
1 kV/mm, 5 kV/mm and 10 kV/mm, are considered. The probability diagram is 
shown in Fig.~\ref{fig1}a, while the random
field distribution in the sample for the case $\Delta$ = 5 kV/mm is shown in
Fig.~\ref{fig1}b. It can be seen that although $\Delta$ is 5 kV/mm, most 
elements has the 
random field below 5 kV/mm. The maximum value can reach as high as 10 kV/mm.

\begin{table}[!hbp]
\centering
\begin{tabular}{lllr}
\toprule
Parameter &Signal    & Value & Unit \\
\midrule
Elastic stiffness moduli&$C_{11}$      
&$2.11\times10^{11}$&N/m\textsuperscript2\\
                        &$C_{12}$      
&$1.07\times10^{11}$&N/m\textsuperscript2\\
                        &$C_{33}$      
&$5.62\times10^{10}$&N/m\textsuperscript2\\
Piezoelectric moduli    &$d_{31}$      &$-3.88$&C/m\textsuperscript2\\
                        &$d_{33}$      &$5.48$&C/m\textsuperscript2\\
                        &$d_{51}$      &$32.6$&C/m\textsuperscript2\\
Dielectric permittivity  &$k_{11}$&$1.75\times10^{-7}$&F/m\\
                        &$k_{22}$&$1.75\times10^{-7}$&F/m\\
Domain wall width       &$\lambda$      &$1$&nm\\
Domain wall energy      
&$G$&$0.7\times10^{-3}$&Jm\textsuperscript2/C\textsuperscript2\\
Mobility    &$\beta$      &$7\times10^{-6}$&A/Vm\\
Eigenpolarization       &$P_{0}$      &0.31&C/m\textsuperscript2\\
Eigenstrain             &$\varepsilon_{0}$&0.262& $\%$\\
\bottomrule
\end{tabular}
\caption{Parameters used in the simulations.}
\end{table}

\begin{figure}[!hbp]
\centering
\includegraphics[width=0.8\textwidth]{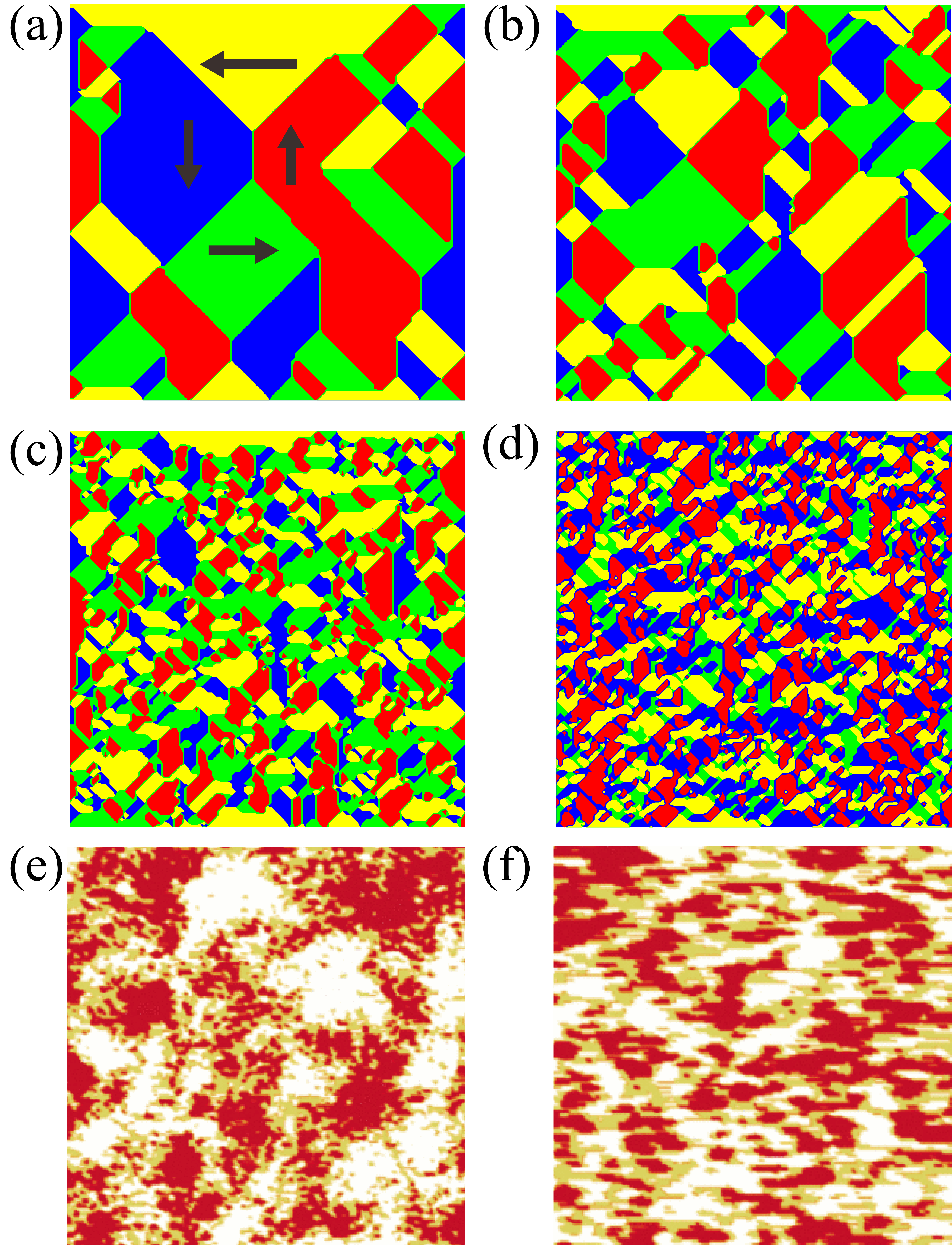}
\caption{ (a-d) Equilibrium domain configuration for four cases with different 
Gaussian distribution variances:  
(a) $\Delta$ = 0 kV/mm;  (b) $\Delta$ = 1 kV/mm; 
(c) $\Delta$ = 5 kV/mm; (d) $\Delta$ = 10 
kV/mm. Four colors represent four polarization orientations of 
domains: yellow-left; green-right; red-top and blue-down. PFM 
images observed on a c-cut SBN single crystals with various compositions:  
(e) SBN61 and (f) SBN75. The PFM images are taken from the paper by 
\cite{shvartsman2008nanopolar}.}
\label{fig2}
\end{figure}  
The equilibrium domain configuration for different values of $\Delta$ can be 
found in 
Fig.~\ref{fig2}.  In 
Fig.~\ref{fig2}a-d four 
colors represent four polarization orientations, respectively. Both 
$90^{\circ}$ and 
$180^{\circ}$ domain walls are visible. Due to the free-force boundary 
condition, 
polarization at the edge of the body tends to point parallel to the boundary. 
By comparing Figs.~\ref{fig2}a-d, it can be concluded that 
the domain size decreases with increasing $\Delta$.  
Figure \ref{fig2}a is for the case $\Delta$ = 0, namely for a conventional 
ferroelectric without random field. As random 
field strength is increased, the material behavior becomes more relaxor-like. 
When $\Delta$ is increased to 5 kV/mm, the domain 
distribution becomes much randomized, as shown in Fig.~\ref{fig2}d. The size of 
the domain becomes smaller 
and the shape becomes more 
twisted.

The domain configuration can be compared with experimental results. 
In the work by \cite{shvartsman2008nanopolar}, the surface polarization status 
of commercial relaxor material 
Sr\textsubscript{x}Ba\textsubscript{1-x}Nb\textsubscript{2}O\textsubscript{6} 
(SBNx) at 293 K was studied by Piezoresponse Force 
Microscopy (PFM). The vertical PFM characterization mode shows the polar 
structures at the surface of the sample.
Fig.~\ref{fig2}e and \ref{fig2}f are two PFM images, which demonstrate 
different domain structures on the surfaces of the sample SBN61 and SBN75, 
respectively. Thereby, tri-modal color code is used: white stands for up, red 
for down, and
yellow for in-plane. It can be seen that for higher Sr content, the domain size 
becomes smaller, and the domain boundaries jag more. Comparison 
between the simulated equilibrium domain structures with the PFM images shows 
that, Fig.~\ref{fig2}b is close to Fig.~\ref{fig2}e, while Fig.~\ref{fig2}c 
appears similar to Fig.~\ref{fig2}f. It can thus be suggested that the Sr 
content is positively correlated to the strength of the random field.

\subsection{Ferroelectric response}
The influence of random field on the macroscopic behavior of relaxor 
ferroelectrics
can be characterized by polarization and strain hysteresis. Here 6 kV/mm 
bipolar loading is adopted for simulation. Gaussian 
distribution variance of random field $\Delta$ 
varies from 0 kV/mm to 50 kV/mm.  

\begin{figure}[!hbp]
\centering
\includegraphics[width=\textwidth]{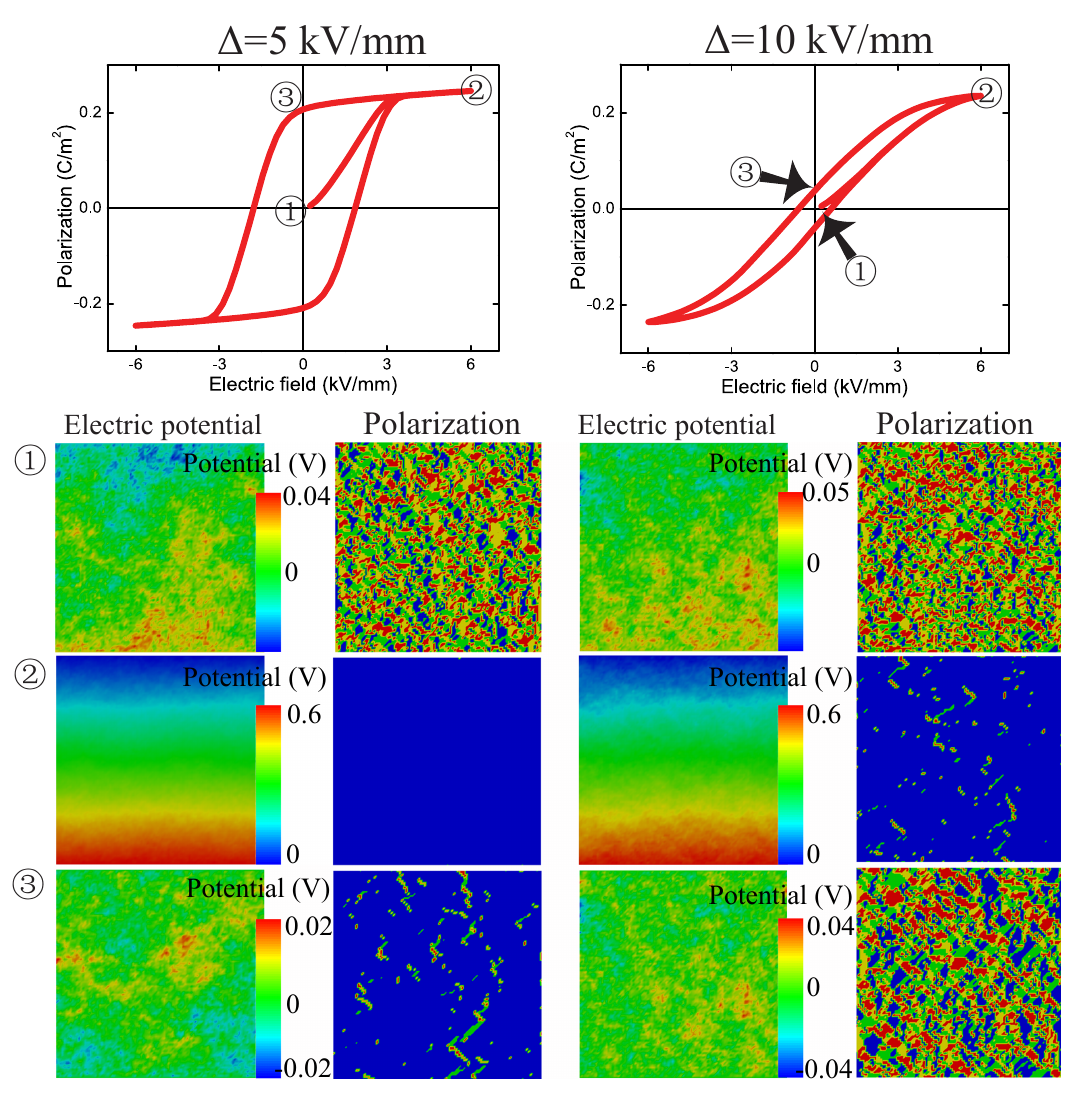}
\caption{Dielectric hysteresis, snapshots of domain structure and electrical 
potential distribution for two different cases: $\Delta$ = 5 kV/mm (left) and 
$\Delta$ = 10 kV/mm (right).}
\label{fig3}
\end{figure}

In order to reveal the influence of the random field, dielectric hysteresis for 
the two cases $\Delta$ = 5 kV/mm and 10 kV/mm are compared in Fig.~\ref{fig3}. 
For small $\Delta $ = 5 kV/mm, the hysteresis has a rectangular shaped 
hysteresis, which is typical for a ferroelectric single crystal. Whereas, for 
large $\Delta$ = 10 kV/mm, the hysteresis becomes very slim. Particularly, the 
 remnant polarization $P_r$ almost vanishes. In order to examine this decrease 
of 
$P_r$, snapshots of domain structure and electric potential distribution during 
the poling are compared in Fig.~\ref{fig3}. Starting from initially random 
distribution (Point 1), the polarization is fully poled at 
maximum field state (Point 2) for both cases. It can be explained by the fact 
that for both cases the applied electric field is stronger than the local 
fields at most sites. 
However, the domain structures at the  remnant state (Point 3) are rather 
different from each other. 
When the applied field 
decreases, the influence of random field becomes 
obvious. Specifically, for
$\Delta$ = 5 kV/mm, the polarization distribution remains almost unchanged, 
because  
the potential barrier prevents the domain switching. However, if the local 
random field is high enough, e.g. $\Delta$ = 10 kV/mm, this barrier can be 
overcome. Thus the polarization distribution becomes randomized again, which 
explains the trivial  remnant polarization.

\begin{figure}[!hbp]
\centering
\includegraphics[width=\textwidth]{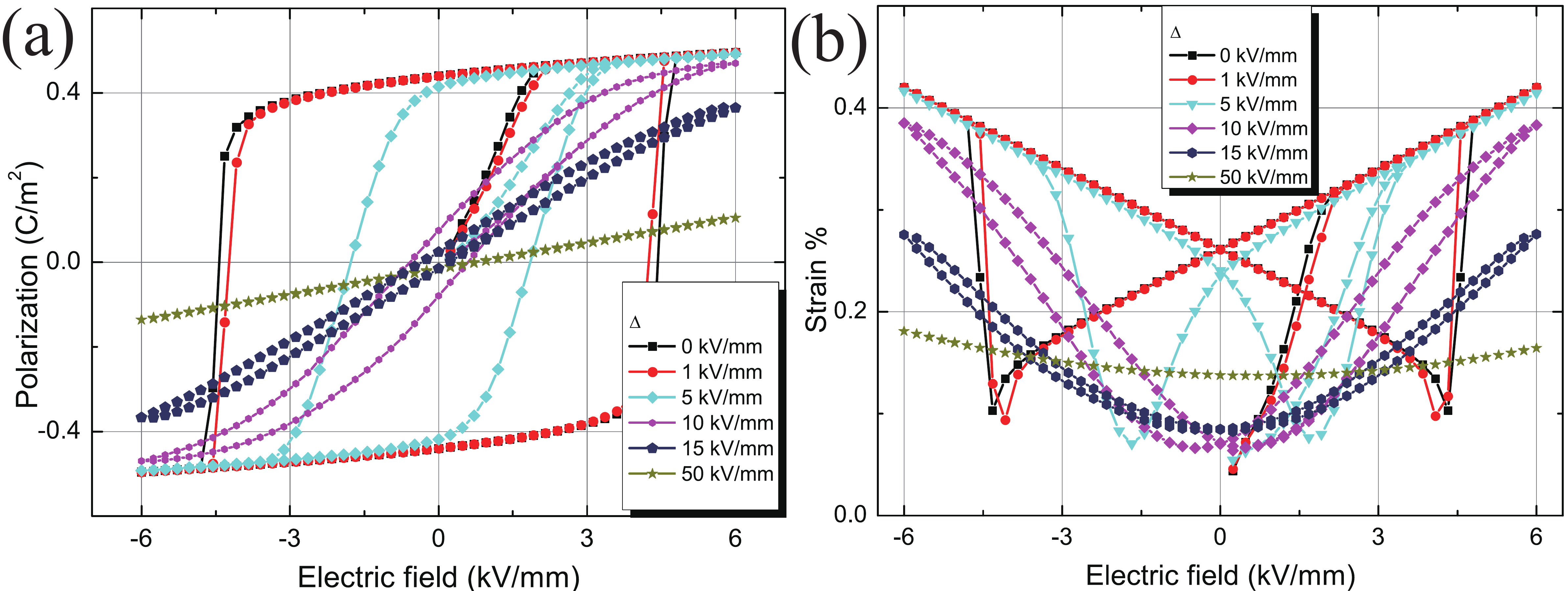}
\caption{(a) Polarization hysteresis and (b) strain hysteresis under bipolar 
loading 
for different cases of random field variance.
}
\label{fig4}
\end{figure}

In Fig.~\ref{fig4} the influence of the random fields on the dielectric and 
strain hysteresis is demonstrated in more details.  
The calculated hysteresis can be classified into three types. 
If $\Delta$ is less than 5 kV/mm, 
the  remnant polarization (P\textsubscript{r}) is close to the maximum 
polarization, while the 
coercive field (E\textsubscript{c}) has a relatively high value of 4.5 kV/mm. 
These two phenomena 
are commonly found in conventional ferroelectric materials. 
Moreover, when external field reaches the 
critical value for the domain switching, the polarization orientation 
changed simultaneously, which leads to a steep slope at the switching point. 
This 
is the typical feature for single crystal~\citep{park1997ultrahigh}.
After domain 
switching process, both  polarization and strain  is in proportion to the 
external field. 

For the sample with moderate random field, e.g. $\Delta$ = 10 kV/mm and 
$\Delta$ 
= 15 
kV/mm, P\textsubscript{r} decreases with the increase of random field. The 
hysteresis 
feature is less obvious, compared with the cases with lower random field. Due 
to 
the lose of hysteresis, the  remnant stain becomes smaller, and the difference 
between maximum stain and  remnant stain becomes larger. Hence, the existence 
of 
random field makes the high-field piezoelectric coefficient (maximum strain 
divided by maximum field) larger. Furthermore, strain changes almost 
quadratically with the applied field, which indicates a typical 
electrostrictive response.

For the sample with very larger random field, e.g. 50 kV/mm, the material 
shows a dielectric feature and exhibit a nearly linear response to the external 
field. The 
random field in this case is so high that external field has little influence 
on the polarization distribution. The polarization direction is determined 
by the local random field distribution, and stays unchanged during the loading. 

From the discussion above, a short conclusion can be drawn that with the 
increase of random field, P\textsubscript{r} decreases, 
the hysteresis effect vanishes gradually, and the material behavior changes 
from 
conventional ferroelectric to relaxor type.

\subsection{Ferroelastic response}               
\begin{figure}[!hbp]
\centering
\includegraphics[width=1\textwidth]{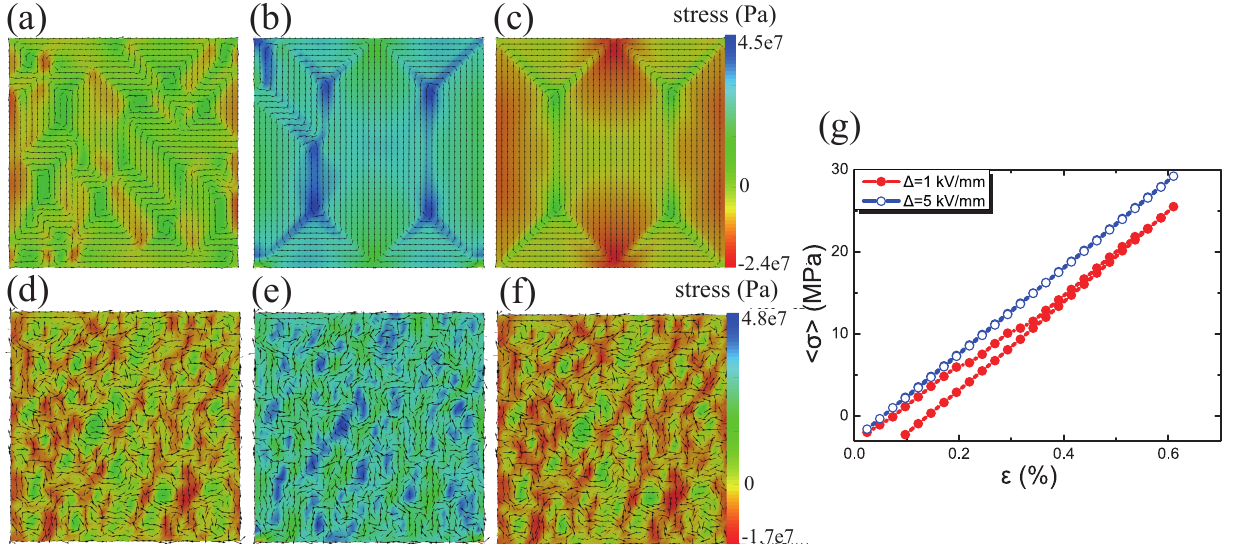}
\caption{(a-f)Domain structure evolution under uniaxial tensile 
loading. (g) Mechanical stress v.s. strain curves.
}
\label{fig5}
\end{figure}

In this section the domain switching induced by pure mechanical loading is 
studied, and the effect of the random field on this response is revealed. 
Figure~\ref{fig5} shows the evolution of the domain structure and of the 
stress distribution in a sample under linearly increased stretch up to $0.6\%$ 
in the vertical direction.  
Two cases of random fields $\Delta$ = 1 kV/mm and 5 kV/mm are considered. 
Charge-free boundary conditions are prescribed.

In the case of low random field $\Delta$ = 1 kV/mm, the initially random 
distributed polarization vectors tend to align along the stretching 
direction, as it is demonstrated in Fig.~\ref{fig5}a-c. Domains with vertical 
polarization become larger. Due to the constrain of charge-free boundary 
conditions, domain structure with multi vertices is formed at the end. This 
mechanical-induced domain switching can also be identified by  
$\sigma$-$\varepsilon$ curve shown in Fig.\ref{fig5}g. The slope of the 
curve decreases during the stretching loading, which indicates that the
domain switching happens in this range. The slope then goes back to its 
original value after domain switching  
process is accomplished. It can be explained by the assumption in the model 
that 
the stiffness tensor is independent of polarization state. During the unloading 
process, the slope of the  
$\sigma$-$\varepsilon$ curve remains constant. Hence the  remnant strain 
emerges, with the value of about 0.1\%. This result is close to the 
experimental measurement on the commercial ferroelectric ceramic PIC 
151~\citep{alatsathianos2000experimentelle}.

\begin{figure}[!hbp]
\centering
\includegraphics[width=1\textwidth]{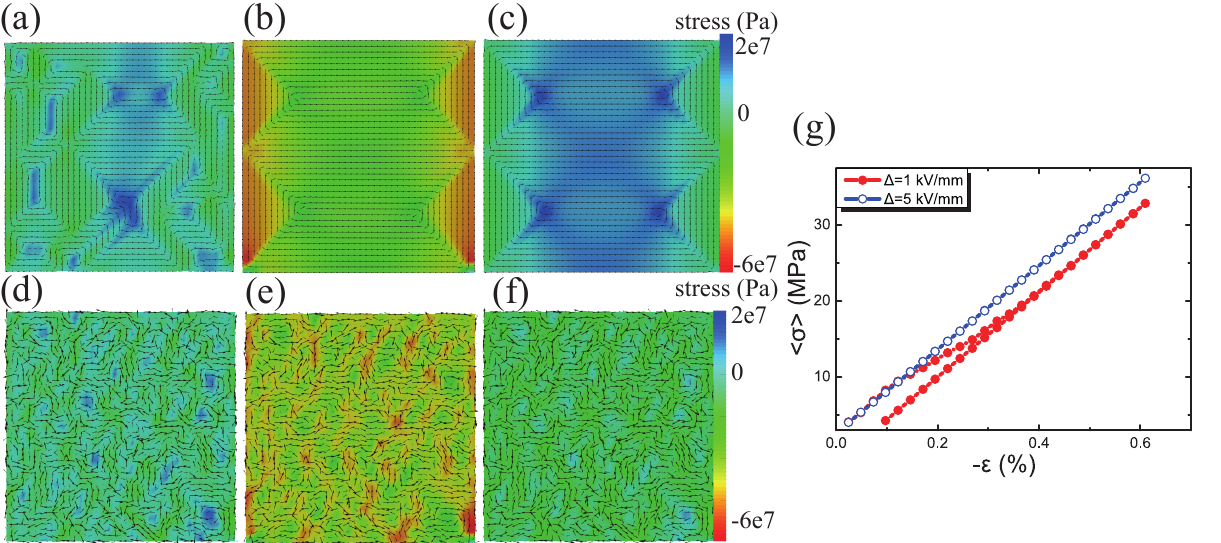}
\caption{(a-f)Domain structure evolution under uniaxial compression. 
(g) Mechanical stress v.s. strain curves.
}
\label{fig6}
\end{figure}

In the sample with high random field $\Delta$ = 5 kV/mm, the coarsening of 
domain is not visible, and the tendency of the vertical polarization is not so 
obvious, as it can be seen in Fig.~\ref{fig5}d-f.
Meanwhile, there is hardly slope change of the $\sigma$-$\varepsilon$ curve 
shown in Fig.~\ref{fig5}g. 
It indicates that there is no apparent mechanical-induced domain switching. The 
random field counteracts the 
influence of mechanical loading on polarization switching. 
Similar phenomenon can be found in the case of compression loading, as it can 
been seen from the simulated domain structures and the stress-strain curve 
shown in 
Fig.~\ref{fig6}.

\subsection{Electromechanical loading}
\begin{figure}[!hbp]
\centering
\includegraphics[width=0.8\textwidth]{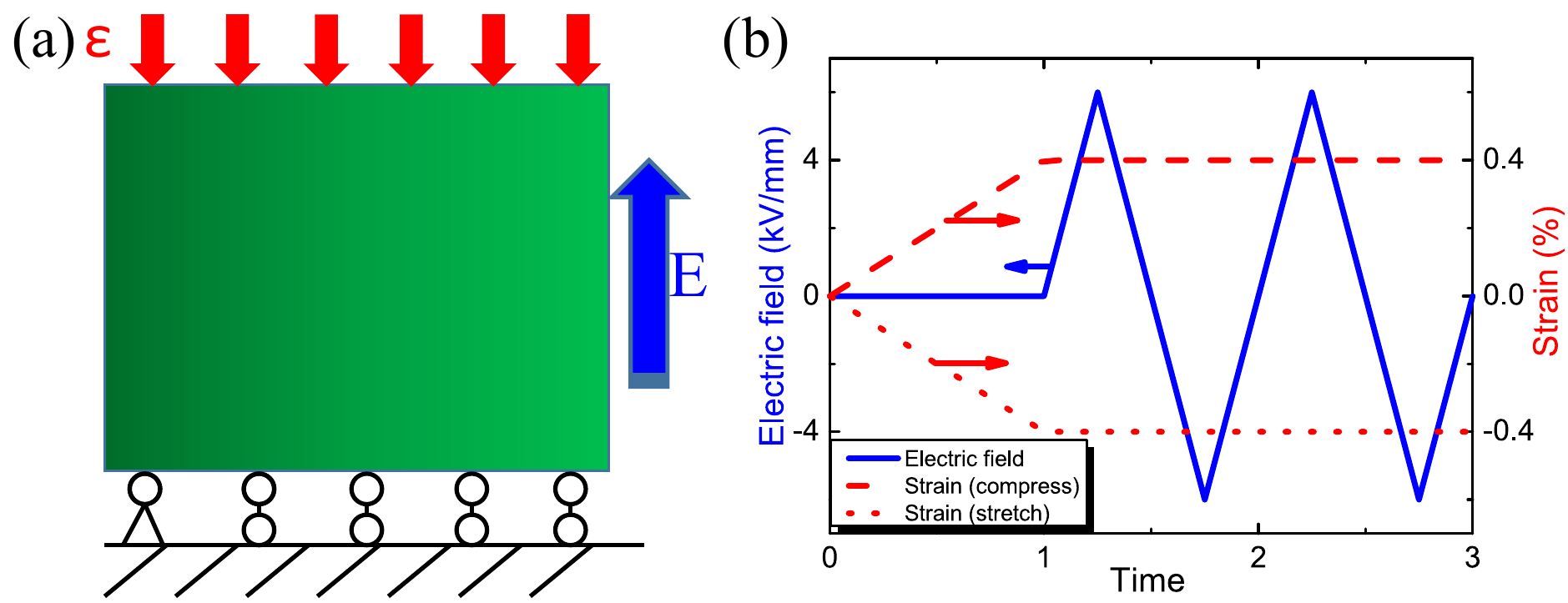}
\caption{(a)Illustration of the simulation setup; (b) Electric and 
mechanical loading history. }
\label{fig7}
\end{figure}

The hysteresis behavior of mechanically assisted bipolar poling on
relaxors is studied. For each of the two random fields with $\Delta$ = 2.3 
kV/mm and 7.5 kV/mm, three loading scenarios are simulated: 1) only bipolar 
electric loading, 2) bipolar electric loading combined with mechanical 
stretching, 3) bipolar electric loading combined with mechanical compression. 
As 
illustrated in Figs.~\ref{fig7}a and \ref{fig7}b, mechanical loading in the 
last 
two scenarios are applied first along the vertical direction. The electric 
bipolar loading is then applied in the same direction, after the 
mechanical 
loading reached a constant. 
The maximum compression/stretch is 0.6\%, and the amplitude of the electric 
field is 6 kV/mm. 

\begin{figure}[!hbp]
\centering
\includegraphics[width=1\textwidth]{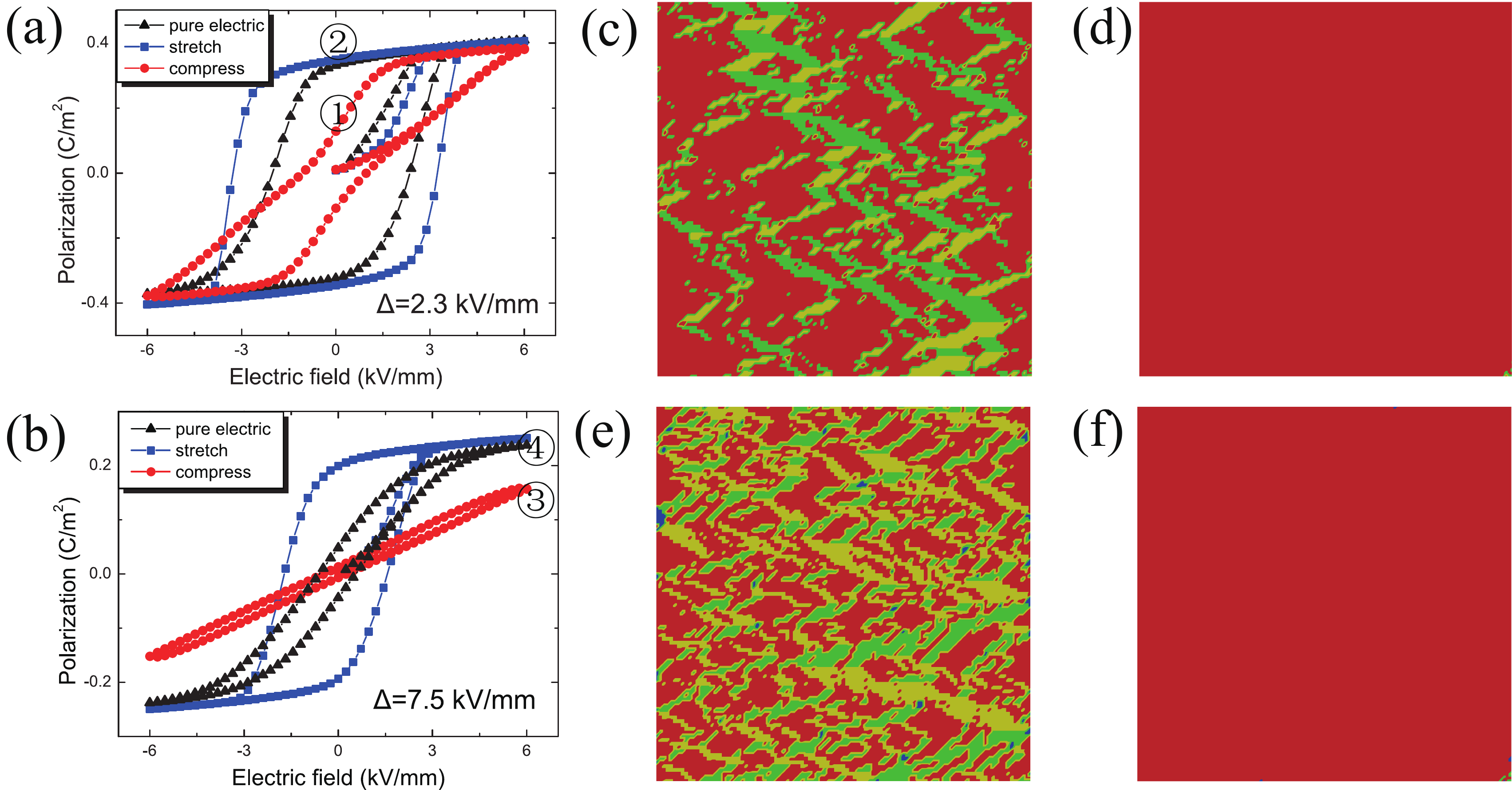}
\caption{Dielectric hysteresis for three loading scenarios for (a) the random 
field with $\Delta$ = 2.3 kV/mm and (b) the random field with $\Delta$ = 7.5 
kV/mm. (c) Domain structure at the  remnant state of the compressed sample 
(Point 
1). (d) Domain structure at the  remnant state Point 2 of the stretched sample 
(Point 2). (e) Domain structure at the maximum field of the compressed sample 
(Point 3). (f) Domain structure at the maximum field of the stretched sample 
(Point 4).}
\label{fig8}
\end{figure}

The three hysteresis of the different loading scenarios shown in 
Fig.~\ref{fig8}a
are for the random field case $\Delta$ = 2.3 kV/mm. For this 
relatively low random field, the polarization at 
maximum field saturates at all the three loading scenarios, and the 
polarization is fully poled. 
However, the  remnant states of the compressed and the stretched sample, 
marked 
as point 1 and point 2 in Fig.~\ref{fig8}a,  
are rather different, after the external field is removed. It can be seen that 
the  remnant polarization of the compressed sample is much less than that of 
the 
stretched one. This feature can be well explained by the domain structure at 
the 
corresponding  remnant states. The domain structure at the  remnant state of 
the 
compressed sample is shown in Fig.~\ref{fig8}c, while that of the stretched 
sample in Fig.~\ref{fig8}d. It is clear that due to the compressive loading, 
part of the polarization is switched to the lateral directions, which leads to 
the drop of the  remnant polarization, whereas in the stretched loading, the 
poled state is maintained.   
Moreover, E\textsubscript{c} of the compressed sample is 
much smaller than that of the stretched one. It can be 
simply explained by the switching energy criterion~
\citep{hwang1995ferroelectric}. The positive mechanical work reduces the work 
needed from external field for switching. In other words, more energy is needed 
for 
compressed sample than for the stretched sample to switch the polarization. 
Similar results can be found in the work by \cite{soh2006phase} on the 
electromechanical 
coupling problems of ferroelectrics. Their simulation results show that both 
E\textsubscript{c} and 
P\textsubscript{r} of compressed samples decrease.

In the case of a rather high random field $\Delta$ = 7.5 kV/mm, different 
features in these hysteresis are observed, as shown in Fig.~\ref{fig8}b. The 
polarization of the compressed sample at the maximum 
electric field does not reach the saturation, because the imprint effect of the 
random field and the effect of the compression loading is strong enough to 
prevent part of the dipoles switching to the vertical direction. On the 
other 
hand, in the stretched sample, the driving force induced by the applied field 
and by the tensile loading overruns the imprint effect of the random electric 
field, and thus the polarization is fully poled. The domain structures at this 
state of the stretched and the compressed sample, shown in 
Fig.~\ref{fig8}e and Fig.~\ref{fig8}f, verify this argument.  
Moreover, not only the compressed sample, but also the free sample has a 
decreased  re polarization.
It implies that the imprint effect of the random field is strong enough, so 
that part of the dipoles are 
switched back to the lateral directions. In the stretched sample, the imprint 
effect of random field is counteracted by the mechanical loading. It explains 
the large  remnant polarization.

In summary, in relaxors the response of mechanically assisted poling can be 
rather complicated, since the imprint effect of the random field plays also an 
important role in the domain switching process, in addition to the mechanical 
loading and the applied electric field. Meanwhile, it can be suggested from 
these results that if large hysteresis is desired for application, the sample 
should be stretched along the direction of the
applied field. On the other hand, if larger piezoelectric or inversed 
piezoelectric  
coefficients are desired, the sample is better to be compressed along the field 
direction.

\section{Concluding remarks}
To conclude, a continuum phase-field model for relaxor 
ferroelectrics is constructed to simulate the domain structure evolution in 
relaxor ferroelectrics. Local random fields are introduced according to the 
random 
field theory, which can appropriately reflect the influence of chemical 
disorder 
in relaxors.
 The strength of the random field is controlled by the variance of the Gaussian 
distribution.
 The simulation results presented in section 4 and the comparison with 
experimental measurements demonstrate that this model can 
reproduce typical relaxor features, such as domain miniaturization, 
small  remnant polarization and large piezoelectric
response.

Under pure electric loading, domain size decreases and domain configuration is 
more 
twisted, as the random field becomes stronger. The  remnant polarization 
decreases with the increase of  random fields.
Under pure mechanical loading, domains size also become smaller as random 
field 
increases, and the strain-induced domain switching can be prevented in the 
presence of larger 
random field.
When bipolar loading is applied on a preloaded sample, the macroscopic 
properties such as  remnant and maximum polarization 
can be modified. These results are in consistent with 
previous study on ferroelectric materials.

This 
model can be further applied, e.g. to study different relaxor materials by 
justifying
parameters, particularly the variance of Gaussian distribution of random field 
($\Delta$). Moreover, if combined with existing phase-field ferroelectric 
models, relaxor/ferroelectric composites can be simulated. Owing to 
the advantage of 
finite element method, structures with complex shapes can be presented.

\section*{Acknowledgements}
The financial support by the 'Excellence Initiative' of the 
German Federal and State Governments and the Graduate School of Computational 
Engineering at Technische Universit\"at Darmstadt is appreciated. The funding by 
the German Science Foundation on the project (Xu121/1-2) in the framework of DFG 
SPP1599 is acknowledged. The discussion with M.S. 
Yang-Bin Ma in the same research group is also
greatly appreciated. Useful discussion in the  group seminar of Prof. J\"urgen 
R\"odel is acknowledged.

\newpage
\section*{Appendix}
The specific forms of the vectors and matrices in section \ref{2d} are given as follows:
\[
\underline{\bm u}
=
\begin{bmatrix}
    u_1 & u_2
\end{bmatrix}^{T}, 
\underline{\bm P}
=
\begin{bmatrix}
   P_1 & P_2
\end{bmatrix}^{T},
\underline{\bm E}
=
\begin{bmatrix}
   E_1 & E_2
\end{bmatrix}^{T},
\underline{\bm D}
=
\begin{bmatrix}
   D_1 & D_2
\end{bmatrix}^{T},
\underline{\bm n}
=
\begin{bmatrix}
   n_1 & n_2
\end{bmatrix}^{T}
=
\begin{bmatrix}
   \frac{P_1}{\sqrt{P_1^2+P_2^2}} & \frac{P_2}{\sqrt{P_1^2+P_2^2}}
\end{bmatrix}^{T},
\]
\[
\underline{\bm \varepsilon}
=
\begin{bmatrix}
   {\varepsilon}_{1} \\ {\varepsilon}_{2} \\ {\varepsilon}_{3}
\end{bmatrix}
=
\begin{bmatrix}
    u_{1,1}\\ 
   u_{2,2} \\ 
0.5(u_{1,2}+u_{2,1})
\end{bmatrix},
\underline{\bm \varepsilon^0}
=
\begin{bmatrix}
   {\varepsilon}_{1}^0 \\ {\varepsilon}_{2}^0 \\ {\varepsilon}_{3}^0
\end{bmatrix}
=
\frac{3 \varepsilon_{0}\sqrt{P_1^2+P_2^2}}{2P_{0}}
\begin{bmatrix}
   n_1^2-\frac{1}{3} \\ n_2^2 -\frac{1}{3}\\ 2n_1n_2
\end{bmatrix},
\underline{\bm \varepsilon^e}
=
\begin{bmatrix}
   {\varepsilon}_{1}-{\varepsilon}_{1}^0 \\ 
{\varepsilon}_{2}-{\varepsilon}_{2}^0 \\ {\varepsilon}_{3}-{\varepsilon}_{3}^0
\end{bmatrix},
\]
\[
\underline{\bm \sigma}
=
\begin{bmatrix}
   {\sigma}_{1}   \\
   {\sigma}_{2}   \\
   {\sigma}_{3}
\end{bmatrix},
\underline{\bm C}
=
\begin{bmatrix}
  C_{11} &C_{12} &0\\
  C_{12} & C_{22} &0\\
  0 & 0 &C_{33}
\end{bmatrix},
\underline{\bm k}
=
\begin{bmatrix}
k_{11}& 0 \\
0 &k_{22}
\end{bmatrix},
\underline{\bm b}
=
\frac{\sqrt{P_1^2+P_2^2}}{P_{0}}\begin{bmatrix}
b_{111}
&b_{122} &b_{112}\\
b_{211} &b_{222} &b_{212}
\end{bmatrix}
\]

Here, subscripts 1 and 2 represents x and y directions in 2-dimensional plane, 
and  Voigt notation is utilized. 
The specific forms of the deformation matrices are given below.
\[
{\underline{\bm B}}_u^I
=
\begin{bmatrix}
N_{,1}^I& 0 \\
0 &N_{,2}^I\\
N_{,2}^I&N_{,1}^I
\end{bmatrix},
{\underline{\bm B}}_{\phi}^I
=
\begin{bmatrix}
N_{,1}^I \\
N_{,2}^I
\end{bmatrix},
{\underline{\bm B}}_{\nabla}^I
=
\begin{bmatrix}
N_{,1}^I&0 \\
0&N_{,2}^I\\
N_{,2}^I&0 \\
0&N_{,1}^I\\
\end{bmatrix},
\]
where $N(x, y)^I$ is the shape function of node I. 
The final  iteration matrix S have the form of :
\[
\underline {\bm S}
=
\begin{bmatrix}
{\bm S}^{11}& {\bm S}^{12}&{\bm S}^{13}& {\bm S}^{14} \\
{\bm S}^{21}& {\bm S}^{22}&{\bm S}^{23}& {\bm S}^{24} \\
{\bm S}^{31}& {\bm S}^{32}&{\bm S}^{33}& {\bm S}^{34} \\
{\bm S}^{41}& {\bm S}^{42}&{\bm S}^{43}& {\bm S}^{44} 
\end{bmatrix}.
\]
In each component in $\underline {\boldsymbol S}^{IJ}$,
\[
\underline {\boldsymbol S}^{IJ}
=
\begin{bmatrix}
{\underline {\boldsymbol{S}}}^{IJ}_{uu}&
 {\underline {\boldsymbol{S}}}^{IJ}_{u\phi}&
 {\underline {\boldsymbol{S}}}^{IJ}_{uP}\\
 {\underline {\boldsymbol{S}}}^{IJ}_{\phi u}&
 { {{S}}}^{IJ}_{\phi \phi}&
 {\underline {\boldsymbol{S}}}^{IJ}_{\phi P}\\
 {\underline {\boldsymbol{S}}}^{IJ}_{Pu}&
 {\underline {\boldsymbol{S}}}^{IJ}_{P\phi}&
 {\underline {\boldsymbol{S}}}^{IJ}_{PP}
\end{bmatrix}
=
\begin{bmatrix}
{\underline {\boldsymbol{K}}}^{IJ}_{uu}&
 {\underline {\boldsymbol{K}}}^{IJ}_{u\phi}&
 {\underline {\boldsymbol{K}}}^{IJ}_{uP}\\
 {\underline {\boldsymbol{K}}}^{IJ}_{\phi u}&
 { {{K}}}^{IJ}_{\phi \phi}&
 {\underline {\boldsymbol{K}}}^{IJ}_{\phi P}\\
 {\underline {\boldsymbol{K}}}^{IJ}_{Pu}&
 {\underline {\boldsymbol{K}}}^{IJ}_{P\phi}&
 {\underline {\boldsymbol{K}}}^{IJ}_{PP}+\frac{1}{\tau}{\underline {\boldsymbol{D}}}^{IJ}_{PP}
\end{bmatrix},
\]
where the tangent matrices are given by
\begin{equation*}
\begin{split}
{\underline {\boldsymbol{S}}}^{IJ}_{uu}=&
\int_{\mathcal S^e} {\underline {\bm B}^{I}_{u}}^{T}{\underline{\boldsymbol{C 
}}}\,\underline {\bm B}^{J}_{u}ds
\\
{\underline {\boldsymbol{S}}}^{IJ}_{u\phi}=&
\int_{\mathcal S^e} {\underline {\bm 
B}^{I}_{u}}^{T}{\underline{\boldsymbol{b}}}^T\underline {\bm B}^{J}_{\phi}ds
\\
{\underline {\boldsymbol{S}}}^{IJ}_{uP}=&
-\int_{\mathcal S^e} {\underline {\bm B}^{I}_{u}}^{T}
\left(
\underline{\boldsymbol 
C}\frac{\partial\underline {\bm \varepsilon^0}}{\partial \underline 
{\bm P}^J}+\frac{\partial \underline{\boldsymbol 
b}^T}{\partial \underline{\bm P}^J}
(\underline 
{\bm 
E}^{\phi}+\underline {\bm 
E}^{random})^{T}
\right)
ds
\\
{\underline {\boldsymbol{S}}}^{IJ}_{\phi u}=&
\int_{\mathcal S^e} {\underline 
{\bm B}^{I}_{\phi}}^{T}{\underline{\boldsymbol{b}}}\,\underline {\bm 
B}^{J}_{u}ds
\\
{{\boldsymbol{S}}}^{IJ}_{\phi \phi}=&
-\int_{\mathcal S^e} {\underline 
{\bm B}^{I}_{\phi}}^{T}{\underline{\boldsymbol{k}}}\,\underline {\bm 
B}^{J}_{\phi}ds
\\
{{\boldsymbol{S}}}^{IJ}_{\phi P}=&
\int_{\mathcal S^e} {\underline {\bm B}^{I}_{\phi}}^{T}
\left(
\frac{\partial 
\underline{\boldsymbol{b}}}{\partial 
\underline{\boldsymbol{P}}^J}
\underline{{\bm 
\varepsilon}^e}+{\underline{\boldsymbol{b }}}\frac{\partial 
\underline{{\bm \varepsilon}^e}}{\partial 
\underline{\boldsymbol{P}}^J}+\underline{\boldsymbol{1}}N^{J}
\right)
ds
\\
{\underline {\boldsymbol{S}}}^{IJ}_{Pu}=&
-\int_{\mathcal S^e}N^{I}
\left(
\frac{\partial{\underline {\bm \varepsilon^0}}^T}{\partial \underline 
{\bm P}}\underline{\boldsymbol 
C}+{(\underline 
{\bm 
E}^{\phi}+\underline {\bm 
E}^{random})}
\frac{\partial \underline{\boldsymbol b}}{\partial 
\underline{\bm P}}
\right)
{\underline {\bm B}^{J}_{u}}^{T} ds
\\
{{\boldsymbol{S}}}^{IJ}_{P \phi}=&
\int_{\mathcal S^e} N^{I}
\left(
\underline{{\bm \varepsilon}^e}^T
\frac{\partial 
\underline{\boldsymbol{b}}^T}{\partial 
\underline{\boldsymbol{P}}}+\frac{\partial \underline{{\bm 
\varepsilon}^e}^T}{\partial 
\underline{\boldsymbol{P}}}{\underline{\boldsymbol{b}}}^T+\underline{\boldsymbol 
{1}}
\right)
{\underline {\bm B}^{J}_{\phi}}
ds
\\
{{\boldsymbol{S}}}^{IJ}_{P \phi}  =
-\int_{\mathcal S^e}&
\Bigg\{
\Bigg[
N^I
\left(
\frac{\partial^2{\underline{{\bm \varepsilon}^0}}^{T}}{\partial 
\underline{\boldsymbol P} \partial \underline{\boldsymbol P}^J}-
\frac{\partial{\underline{{\bm \varepsilon}^0}}^{T}}{\partial 
\underline{\boldsymbol P}}
\frac{\partial{\underline{\bm \sigma}}^{T}}{\partial \underline{\boldsymbol 
P}^J}
\right)
-N^I
\left(
\frac{\partial{\underline {{\bm \varepsilon}^e}}^{T}}{\partial 
\underline{\boldsymbol P}^J}
\frac{\partial {\underline{\boldsymbol{b}}}^T}{\partial \underline{\boldsymbol{P}}}+
{{\underline {{\bm \varepsilon}^e}}^{T}}
\frac{\partial^2 {\underline{\boldsymbol{b}}}^T}{\partial 
\underline{\boldsymbol{P}}\partial \underline{\boldsymbol{P}^J}}
\right)
(\underline {\bm E}^{\phi}+\underline {\bm E}^{random})^T 
\\
  \qquad  &+\beta_1N^I\frac{\partial^2 \psi}{\partial 
\underline{\boldsymbol{P}}\partial 
\underline{\boldsymbol{P}^J}}+\beta_2{{\underline{{\boldsymbol{B}}}_P^{I}}}^T{ 
\underline {{\boldsymbol{B}}}_P^{J}}
\Bigg]
-\frac{\beta}{\tau}N^I\underline{\boldsymbol{1}}N^J
\Bigg\}
ds.
\end{split}
\end{equation*}


\newpage

\large
\bibliographystyle{model2-names}
\bibliography{ref.bib}

\end{document}